\documentclass[12pt,preprint]{aastex}

\slugcomment{Not to appear in Nonlearned J., 45.}

\shorttitle{Image Rotation and Subtraction}
\shortauthors{Dou et al.}

\begin{document}


\title{A HIGH-CONTRAST IMAGING ALGORITHM: OPTIMIZED IMAGE ROTATION AND SUBTRACTION}


\author{Jiangpei Dou\altaffilmark{1,2}, Deqing Ren\altaffilmark{3,1,2}, Gang Zhao\altaffilmark{1,2}, Xi Zhang\altaffilmark{1,2}, Rui Chen\altaffilmark{1,2}, and Yongtian Zhu\altaffilmark{1,2}}


\altaffiltext{1}{National Astronomical Observatories/Nanjing Institute of Astronomical Optics \& Technology, Chinese Academy of Sciences, Nanjing 210042, China; jpdou@niaot.ac.cn}
\altaffiltext{2}{Key Laboratory of Astronomical Optics \& Tech
nology, Nanjing Institute of Astronomical Optics \& Technology, Chinese Academy of Sciences, Nanjing 210042, China}
\altaffiltext{3}{Physics \& Astronomy Department, California State University Northridge, 18111 Nordhoff Street, Northridge, California 91330-8268, USA}


\begin{abstract}
Image Rotation and Subtraction (IRS) is a high-contrast imaging technique which can be used to suppress the speckles noise and facilitate the direct detection of exoplanets. IRS is different from Angular Differential Imaging, in which it will subtract a copy of the image with $180 ^\circ$ rotated around its point-spread function (PSF) center, rather than the subtraction of the median of all of the PSF images. Since the planet itself will be rotated to the other side of the PSF, IRS does not suffer from planet self-subtraction. In this paper, we have introduced an optimization algorithm to IRS (OIRS), which can provide an extra contrast gain at small angular separations. The performance of OIRS has been demonstrated with ADI data. We then made a comparison of the signal-to-noise ratio (S/N) achieved by algorithms of locally optimized combination of images and OIRS. Finally we found that OIRS algorithm can deliver a better S/N for small angular separations.
\end{abstract}





\section{INTRODUCTION}

Until now over 1800 exoplanets have been detected mostly by the indirect radial velocity and transiting approaches, yielding several important physical information such as masses and
Until now over 1800 exoplanets have been detected mostly by the indirect radial velocity and transiting approaches, yielding several important physical information such as masses and radii. The study of the physics of planet formation and evolution will focus on giant planets through the direct imaging. However, the direct imaging of exoplanets remains challenging, due to the large flux ratio difference and the nearby angular distance. In practical observations, direct imaging is mainly limited by the bright quasi-static speckles that will contaminate the faint companions (Marois et al. 2005; Masciadri et al. 2005). The speckle noise is mainly caused by the wave front distortion from imperfection optics and the non-common path aberrations.

Currently, most of the imaged exoplanets are very young (1$\sim$100 Myr) and giant ($>3M_J$), with a moderate contrast of only$10^{-4}$$\sim$$10^{-5}$ in the near infrared to mid-infrared wavelength (Marois et al. 2008, 2010; Lagrange 2009, 2010). The moderate contrast makes it possible to be imaged from the ground based on current technique. Among them, most were detected by an adaptive optics with a specialized observing strategies such as Angular Differential Imaging (ADI) (Marois et al. 2006) and corresponding image post-processing methods such as locally optimized combination of images (LOCI) (Lafreni\`{e}re et al. 2007). However, it is difficult to obtain a reference image that is highly correlated with the target image, since the quasi-static speckles vary with time (Marois et al. 2005, 2006). Thus LOCI introduce a way of combining reference point-spread function (PSF) images to optimize the speckle attenuation.

LOCI is applied for ADI data has proven to be one of the most powerful speckle suppression methods for ground-based observations and is now broadly used in high-contrast imaging research. However, it is found that LOCI algorithm suffers from planet self-subtraction, especially at small angular separations (Mawet et al. 2012). For LOCI, a compromise has to be made between the speckle noise correlation and a sufficient companion displacement (Lafreni\`{e}re et al. 2007). A parallactic rotation between the target image and those used as reference will guarantee the planet will not be subtracted. But the minimum parallactic rotation angle will become relatively small for small angular separations in order to get a highest speckle correlation. As a result, the planet itself will be partially subtracted.

Small-angle high-contrast imaging is technically and scientifically appealing because it enables the use of small telescopes and potentially increases the yield of nearby faint companions (Mawet et al. 2012). Image Rotation and Subtraction (IRS) is a high contrast imaging technique and initially proposed to work with an Ex-AO coronagraphic system (Ren et al. 2012 a). Different from ADI, IRS will subtract a copy of the image with $180^\circ$ rotated around its PSF center to attenuate the speckle, rather than the subtraction of of the median of all of the PSF images. Thus the planet will not be subtracted since it has been rotated on the other side of its PSF. In our previous work, we have fully demonstrated that the speckles at any separation on the PSF image will be subtracted uniformly. Therefore, IRS can effectively attenuate the speckles at small angular separations,which allows the use of small telescopes for exoplanets imaging research and meanwhile will potentially increase the yield of planet closer to its host star with large ground-based telescopes. To further improve the performance, here in this paper we firstly introduce an optimization algorithm into IRS (hereafter named as OIRS).

The IRS technique and numerical simulation results are presented in Section 2. In Section 3, both LOCI and OIRS algorithm are reviewed. The application of OIRS algorithm to ADI data and its performance are presented in Section 4. Conclusions and future work are discussed in Section 5.

\section{IRS TECHNIQUE}

\subsection{Principle of IRS}

IRS is a high-contrast imaging technique that is initially proposed to work with Ex-AO coronagraph system, which has the possibility to reduce the speckle noise up to the third order. Since the speckles at any separation on the PSF image will be subtracted uniformly and will not be limited by the field rotation, IRS can effectively attenuate the speckles at small angular separations. Here we will review IRS and briefly introduce its principle.
The speckle noise can be decomposed as different orders in respect to the wave-front error and it provides an opportunity to subtract and thus reduce the speckle noise. The wave-front error or phase $\phi$ of an electromagnetic wave on the pupil can be expressed as a Taylor series $e^{i \phi}=\sum_{n=1}^{\infty}\frac{(i \phi)^{n}}{n!}$. Here we will show that the IRS technique can fully take the advantage of the high contrast provided by the coronagraph, as well as the wave front corrected by the adaptive optics.
For an optical system with wave-front error, the electric field of the electromagnetic wave at the pupil plane can be expressed,
\begin{equation}
E_0=Pe^{i\phi},~~~
\label{eq:LebsequeI}
\end{equation}
where $P$ is the pupil function which can be apodized pupil with a stellar coronagraph or unapodized clear pupil.
The electric field of the electromagnetic wave on the focal plane for an optics system with phase aberration is the Fourier transform of the aberrated electric field on the pupil plane
 \begin{equation}
E_f=\overrightarrow{Pe^{i\phi}}.
\label{eq:LebsequeII}
\end{equation}
Then the distorted PSF intensity distribution can be represented as,
 \begin{equation}
I=|\overrightarrow{Pe^{i\phi}}|^2=\overrightarrow{Pe^{i\phi}}\bullet\overrightarrow{Pe^{i\phi*}}\approx\overrightarrow{P}+I_1+I_2+I_3+I_4,
\label{eq:LebsequeIII}
\end{equation}
where the over arrow denotes 2D Fourier transform and * represents the complex conjugate operation; in the above equation, we use the fourth-order Taylor series to calculate the $e^{i\phi}\approx1+i\phi-\phi^2/2-i\phi^3/6+\phi^4/24$ with higher orders truncated which may generally contribute less to the speckle noise).
Here we assume that the pupil is symmetric for simplicity. In that case, the terms $|\overrightarrow{P}|^2$ and the second order and fourth order of the speckles in the wavefront error ($I_2$ and $I_4$) are even and symmetric around the PSF center. The even-order speckles can be eliminated by carrying out the IRS algorithm, leaving the odd-order speckles, which are anti-symmetric. Here we only provide explicit expressions for the odd-order speckles. The first-order and third-order speckles are given as:
\begin{equation}
I_1=\overrightarrow{A}[2Im(\overrightarrow{\phi A^*})],
I_3=Im[\overrightarrow{\phi A}\bullet (\overrightarrow{\phi^2 A^*})] ,
\label{eq:LebsequeIV}
\end{equation}

where Im represents the imaginary part of a complex.

For a coronagraph that is designed to deliver a contrast of $10^{-5}$, for example, the $\overrightarrow{A}$ will have a value on the order of 3 x $10^{-3}$ in the discovery area, and the first-order speckle will be suppressed 10 times better than that of the unapodized pupil. More details can be found in our paper (Ren et al. 2012a).

Once the static wave-front aberration is corrected with an Ex-AO system, the residual wave-front error $\phi$ will be randomly changed since the incoming atmospheric turbulence induced wave-front error is randomly variable. As a result, the residual odd-order speckles $I_1$ and $I_3$ could, in principle, be averaged out to zero by co-adding infinite short-exposure images, or be significantly reduced by co-adding a large number of short-exposure images.

\subsection{Numerical Simulation of IRS}

The nature of speckle attenuation of IRS at small angular separations has been testified by numerical simulations in our previous work, which has demonstrated an effective speckle attenuation at small angular separations down to  $1-2 \lambda/D$  (Ren et al. 2012 a). To demonstrate that IRS can uniformly remove the speckles and does not suffer from planet self-subtraction, we make a systematic analysis based on numerical simulation.

In the simulation, we similarly generate 100 independent AO-corrected phase maps with a moderate Strehl ratio of 0.57 (Ren et al. 2012 a). We use a coronagraph with a contrast of $10^{-5}$ in the simulation (Ren \& Zhu 2007). Figure 1 shows the co-added PSF of 100 images and the combined PSF after IRS, respectively. Since the speckles at any separation on the PSF image will be subtracted uniformly and will not be limited by the field rotation, IRS can effectively attenuate the speckles at small angular separations. Figure 2 shows the achievable contrasts. IRS has dramatically improved the contrast with a gain of more than 10 times at an angular separation down to $1 \lambda/D$ .

To further demonstrate that IRS is not limited by planet self-subtraction, here we induce several artificial planets in each PSF in an ADI mode. Five planets in total are added on each PSF image at an angular separation between  $2-9 \lambda/D$ . Figure 3 shows the IRS reduced images. All planets after IRS has a sufficient signal-to-noise ratio (S/N), which demonstrates IRS does not suffer from the planet self-subtraction. Due to the IRS procedure, five dark spots occur at the symmetrical position where the planets locate; however, it will not influence the judgment of the astrometry position of the planets, since the dark spots have a negative value and can finally be filtered out.

\section{REVIEW OF LOCI AND OIRS}


The LOCI algorithm, as detailed in Lafreni\`{e}re et al. (2007), is to construct an optimized reference PSF image from a set of reference images. The heart of the algorithm is to build the image as a linear combination of available reference images, and the coefficients of the combination are optimized inside multiple subsections of the image independently to minimize the residual noise within each subsection.

To build a reference image, one has to reach a compromise between the speckle noise correlation and a sufficient companion displacement (Marois et al. 2006). A parallactic rotation between the target image and those used as reference will guarantee the planet will not be subtracted. However, the minimum rotation angle will decrease as the inverse of the angular separation. Therefore, the allowable rotation angle will become relatively small for small angular separations in order to get a highest speckle correlation. As a result, the planet itself will be partially subtracted, especially at small angular separations.

OIRS, as a modification of LOCI, does not reconstruct the reference PSF image from those with a sufficient parallactic rotation. In OIRS, each single image will subtract the $180^\circ$ rotation of the combination of the same image and those neighboring images with a small parallactic rotation. Therefore, those combined as the reference image for OIRS can always have a highest speckle correlation. Since the optimization region includes the subtraction region, OIRS partially suffers from the planet self-subtraction. However, the self-subtraction is relatively moderate.

The goal of the optimization algorithm is to minimize the noise in each subsection after the subtraction of rotated image from the target image, where the coefficients in each subsection are variables to be optimized. The coefficients will be computed by minimizing the sum of the square residuals of the subsection of $O^R$ from $O^T$, shown as follows:

\begin{equation}
 \sigma^{2}=\sum_{i}(O_i^R-O_i^T)^2
\label{eq:LebsequeV}
\end{equation}
where $O^T$ and $O^R$ are the optimization subsection of the target image and the rotation of the target image, respectively; $i$ represents each pixel position in the subsection.

Before performing the OIRS, we have been carefully chosen the area and shape of the subsection as well as the optimization subtractions. The associated parameters such as A: the area, g: ratio of the radial and azimuthal widths of the optimization subsection, are similar as that used by Lafreni\`{e}re et al. (2007). Both concentric annular and spiral pattern are used to define the optimization subsections. Finally, we choose a spiral pattern which has a form of $r=(a\theta)^{b}$ in polar coordinates. The spiral pattern will be an Archimedean spiral when $b=1$; well if $b>1$, the section in the inner part of the image is smaller than those of the outer part. Figure 4 shows an example of the subsection under the spiral pattern: optimization subsections for OIRS (shaded in gray) and the subtraction subsections (bright sectorial regions). The subtraction region is included in the optimization region.

\section{APPLICATION OF OIRS TO ADI DATA}

\subsection{Observation Data and Image Preparation}

The observation data of HR 8799 system is downloaded from the ESO Science Archive. The data is taken in Ks band by instrument of VLT/NaCo working in a pupil tracking mode to allow for ADI. Here in this paper we use the data taken on 2009, October 8, the same data used by Currie et al. (2011). Each piece of image data is composed of hundreds of frames with a 0.3454 s exposure time which is stored in a standard NaCo data cube format. There are totally 163 data cubes with a field rotation of $66.43 ^\circ$.

In the first step, we remove the bad AO corrected frames. A criterion is defined as the image intensity ratio between region within radius of $2 \lambda/D$ and that of $9 \lambda/D$ and will be used to judge the AO performance. Frames will be removed if the criterion is lower than 0.45. Then a shutter-closed dark image having the same observing parameters is first subtracted to remove the detector electronic bias, and then the image is divided by a flat field to normalize the spatial variations in sensitivity. Since the NaCo data has been taken under a four-point dithering observation mode, it should wash out image distortion errors (Currie et al. 2011). And the thermal background can be estimated by median combine all of the images. We use a normalized cross-correlation technique to register the image with the saturated part marked. Then the registered images are combined and centered (the center is calculated by cross-correlation between one PSF image with $180 ^\circ$ rotated copy of the image).

\subsection{Data Reduction}

Recently, our group has developed a data reduction pipeline for infrared high contrast imaging. The pipeline includes basic infrared data reduction (e.g. dark current, flat field, background calibrations), imaging registration, and speckle suppression technique of ADI and IRS as well as the algorithms of OIRS and LOCI. Here ADI is referring to the ADI algorithm used by Marois in 2006.  The pipeline is written in IDL, a well-known and frequently used language in astronomical environments.

To demonstrate its potential performance, we have used both IRS and OIRS to reduce the data. For comparison purpose, we have also used the ADI algorithm by Marois et al. (2006) and LOCI. Figure 5 shows the reduced images by using ADI algorithm (Marois et al. 2006) and IRS, respectively. The outer three exoplanets around HR 8799 are clearly imaged. And IRS provides a better S/N for HR 8799 d. For LOCI reduction, the procedure closely follows that of Lafreni\`{e}re et al. (2007), in which we get a similar reduced image with Currie et al. (2011). The reduced images by LOCI and OIRS are shown in Figure 6, in which all exoplanets has been imaged. OIRS provide a better S/N especially for HR 8799 d and e, the two innermost planets. For LOCI reduced data, both planets d and e appears slightly smaller than b and c due to the planet self-subtraction.

Table 1 lists a summary of the S/N of planets around HR 8799 by using above algorithms, calculated in an annulus region in which the noise is defined as an annulus with a width equal to one FWHM of the PSF. Finally, it is found that the achievable speckle attenuation by IRS and OIRS is better for small angular separations. IRS and OIRS can provide a comparable performance at large angular separations, which may be due to the sufficient field rotation at large angular separations, as expected. Although a little degradation than ADI and LOCI occurred for the outer planets of b, which is acceptable since we are mainly focus on the speckle attenuation at small angles.

\subsection{Further Comparison With LOCI Algorithm}

A further comparison of the OIRS algorithm with LOCI is made to eliminate the potential influence of random noise in this data set. We follow a similar procedure that has been used in Lafreni\`{e}re et al. (2007, see their Section 4.4). Several artificial planets are added on the same data set that is used in Section 4.2, in an ADI mode. In each frame of the data set, planets are inserted at three different azimuth angles. For demonstration purpose, here it shows images with azimuth angles of 15, 90 and 165 $^\circ$, respectively. To be similar to the actual planets radial position and S/N, here eight artificial planets are added along one radius, in which the innermost one has an angular separation of $\sim 0.375 \arcsec$, the same as the HR 8799 e, by steps of $0.15 \arcsec$. The final frame has a field rotation of 66 $^\circ$, which is also the same as the data set in above section. The intensities of the artificial planets are adjusted to yield a S/N of $10 \sigma$ for the three inner most planets and $> 20 \sigma$ for all other planets with the OIRS algorithm. Then both OIRS and LOCI are carried on these images in the data set. To be fairly to evaluate the performance of two algorithms, we have employed the same optimization parameters (same area and shape of the subtraction and optimization subsection). Figure 7 shows the compassion of the residual image between LOCI and OIRS. Most of the signals from the innermost planets by LOCI have been subtracted due to the limitation of planet self-subtraction; however, these planets have been only partially subtracted by OIRS. OIRS yields a better speckle attenuation at small angular separations, since it can subtract speckles uniformly at all angular separations and does not seriously suffer from the planet self-subtraction at small angular separations. At large angular separations, both algorithm provide a sufficient S/N. The achievable contrast by two algorithms is shown in Figure 8, in which OIRS can provide 5-10 times better contrast gain than LOCI at angular separations between $0.3-0.6 \arcsec$. The inner working angle for high-contrast imaging has been reduced twice than previous results, which is a unique feature that other current approaches do not provide, which allows the use of small telescopes for exoplanets imaging research and meanwhile will potentially increase the yield of planets closer to its host star with large ground-based telescopes.




\section{CONCLUSION AND FUTURE WORK}

Small-angle high-contrast imaging is technically and scientifically appealing because it enables the use of small telescopes and potentially increases the yield of nearby faint companions (Mawet et al. 2012). IRS has a nature to attenuate the speckles at small angular separations, because it is not limited by the field rotation and can subtract speckle uniformly. OIRS is optimized algorithm that has been firstly introduced into IRS. The principle of OIRS and its performance has been testified for ADI data. It is found that the OIRS may provide a better speckle attenuation (5-10 times higher) at smaller separations (IWA has reduced twice), making it possible to be used for exoplanets on small telescopes. With current data taken without an Ex-AO coronagraph, we have fully demonstrated in this paper that IRS/OIRS can provide a compatible performance. Since IRS is originally proposed and optimized for an Ex-AO coronagraphic system, better performance is expected in the future observations. We will further testify its performance in our ongoing observations, in which a very compact Ex-AO system (Ren et al. 2012 b, 2014) has been developed and has been recently installed as a visiting instrument on ESO 3.58-m NTT and will be installed on the 3.5-m ARC telescope at Apache Point Observatory in early 2015. And a high-contrast coronagraph that is optimized for ground based telescope with central obstruction and spiders (Ren et al. 2010), will be integrated in the Ex-AO system in the next observation runs. Both NTT and ARC will support ADI mode and FOV rotation compensation mode. The IRS technique will be further testified for observation data without field rotations. We will discuss the result in our future work.




\acknowledgments

We thank the anonymous referee for valuable comments, which significantly improved the manuscript. We also thank Dr. Currie and Thalmann for helpful discussions on the achieve data and ADI. This work was made possible based on data obtained from the ESO Science Archive Facility under request number of jpdou/28278. This work was supported by the NSFC (Grant Nos. 11220101001, 11433007, 11328302, 11373005, and 11303064), the special funding for Young Researcher of Nanjing Institute of Astronomical Optics $\&$ Technology, the special fund for astronomy (Grant No. KT2013-022) of CAS, and a grant from the John Templeton Foundation and National Astronomical Observatories. Part of the work described in this paper was carried out at California State University Northridge, with support from the National Science Foundation under Grant ATM-0841440.

\clearpage


\clearpage

\begin{figure}
\epsscale{1.2}
\plottwo{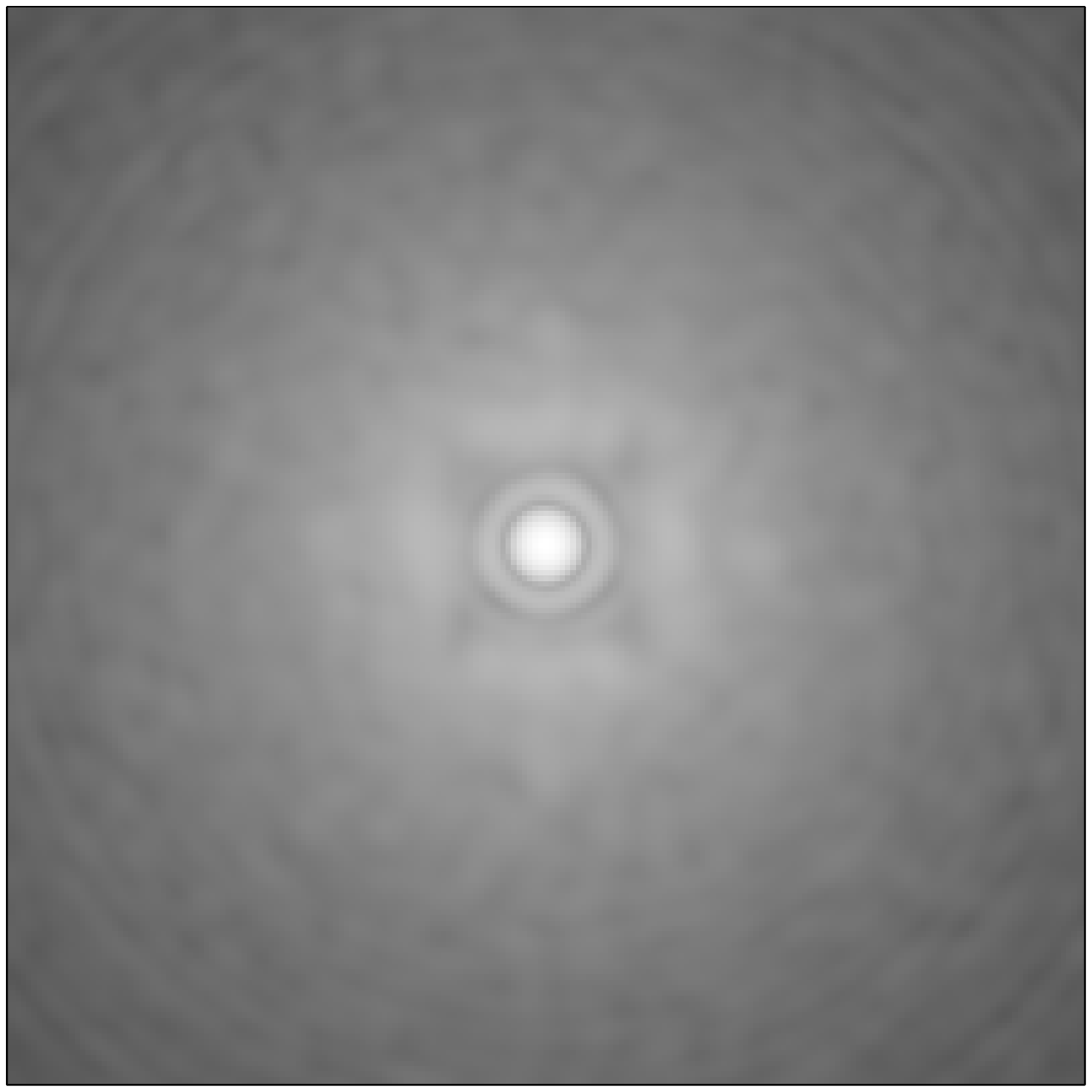}{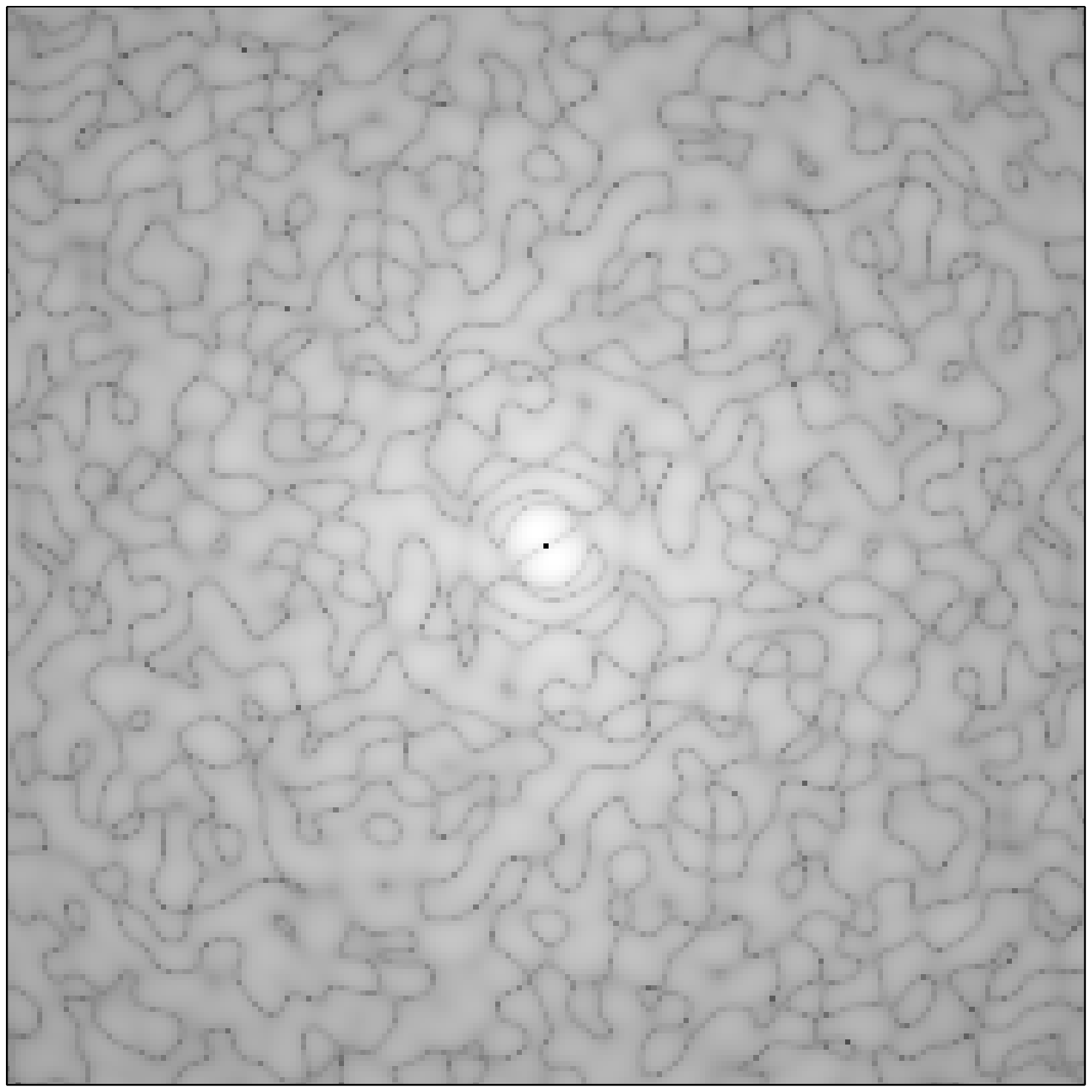}
\caption{PSFs with a $10^{-5}$ contrast coronagraph and a 0.57 Strehl ratio. Left: 100 image combined PSF; right: 100 image residual PSF after IRS. The PSFs are shown in log scale to clearly see that the speckle can be attenuated uniformly at different angular separations.\label{fig1}}
\end{figure}

\begin{figure}
\epsscale{.8}
\plotone{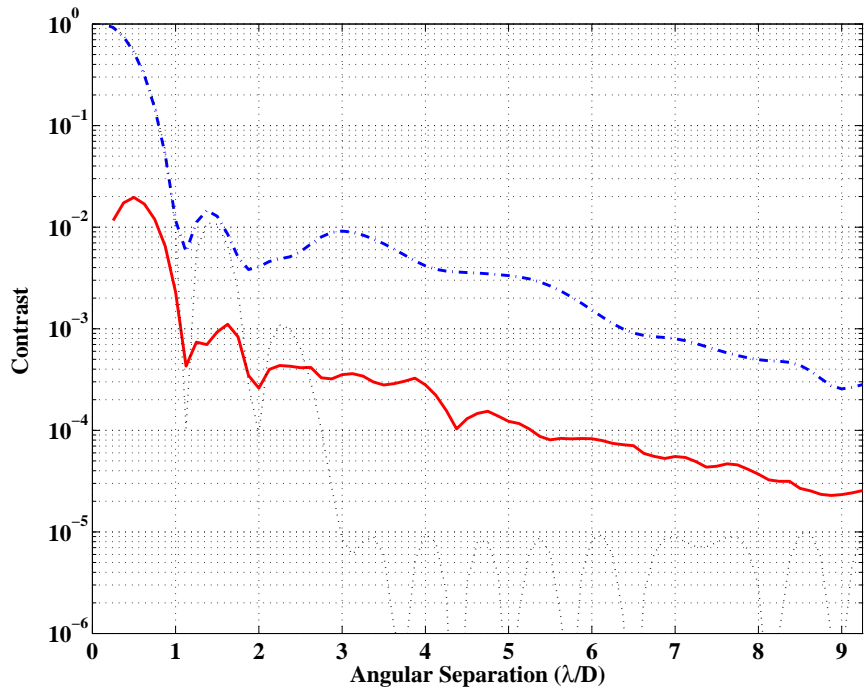}
\caption{Contrasts with a $10^{-5}$ contrast coronagraph and a 0.57 Strehl ratio. Dashed line: theoretical contrast; dash-dotted line: 100 image contrast; solid line: 100 image residual contrast after IRS. See the electronic edition of the journal for a color version of this figure.\label{fig2}}
\end{figure}

\begin{figure}
\epsscale{.8}
\plotone{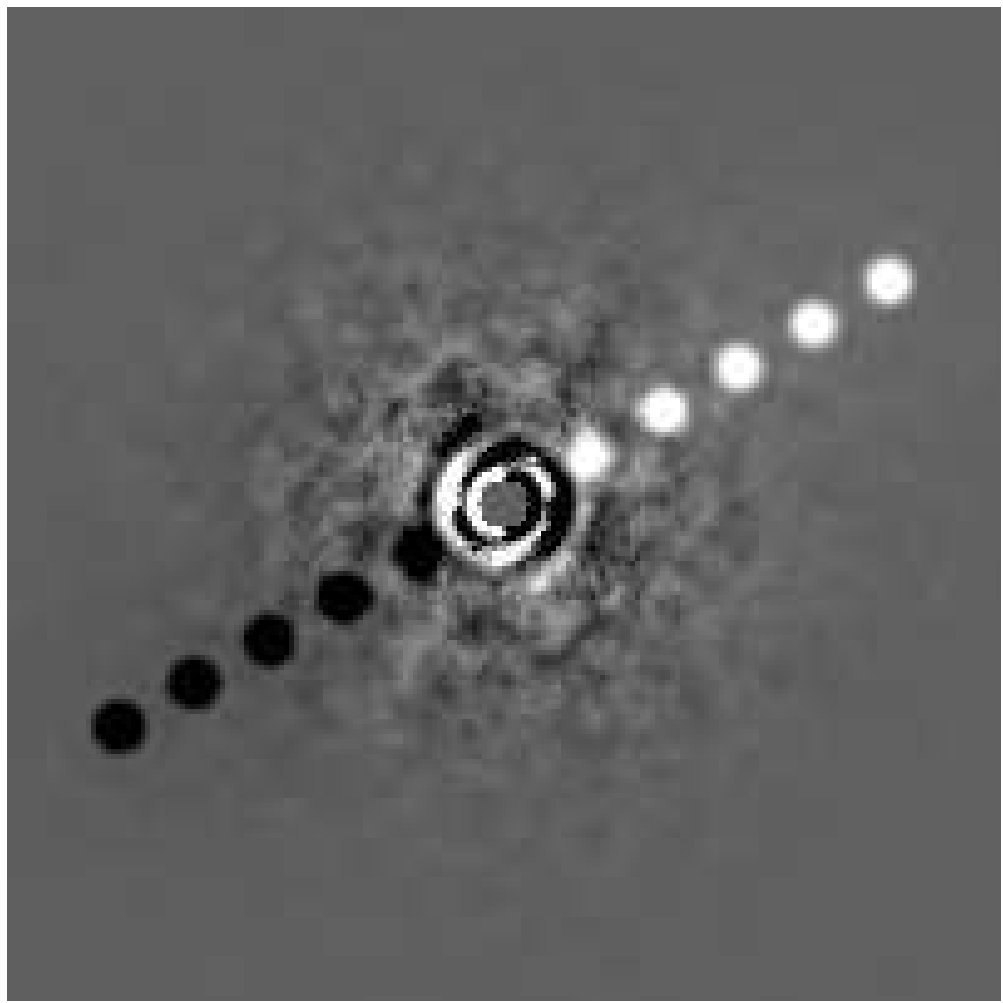}
\caption{IRS reduced data with artificial planets.\label{fig3}}
\end{figure}

\begin{figure}
\epsscale{.8}
\plotone{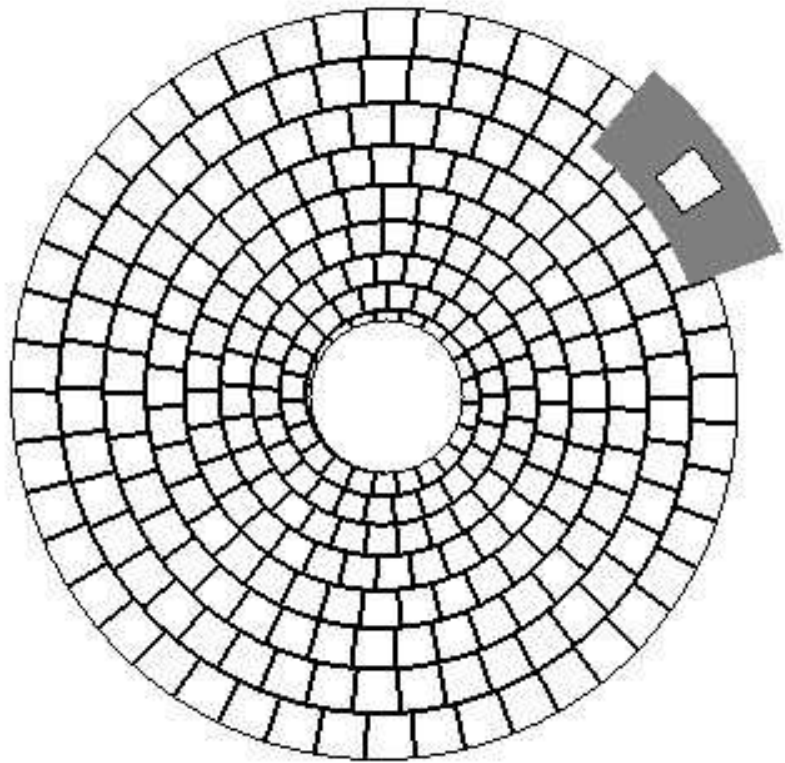}
\caption{An example of the subtracttion and optimization subsection. \label{fig4}}
\end{figure}


\begin{figure}
\epsscale{1.2}
\plottwo{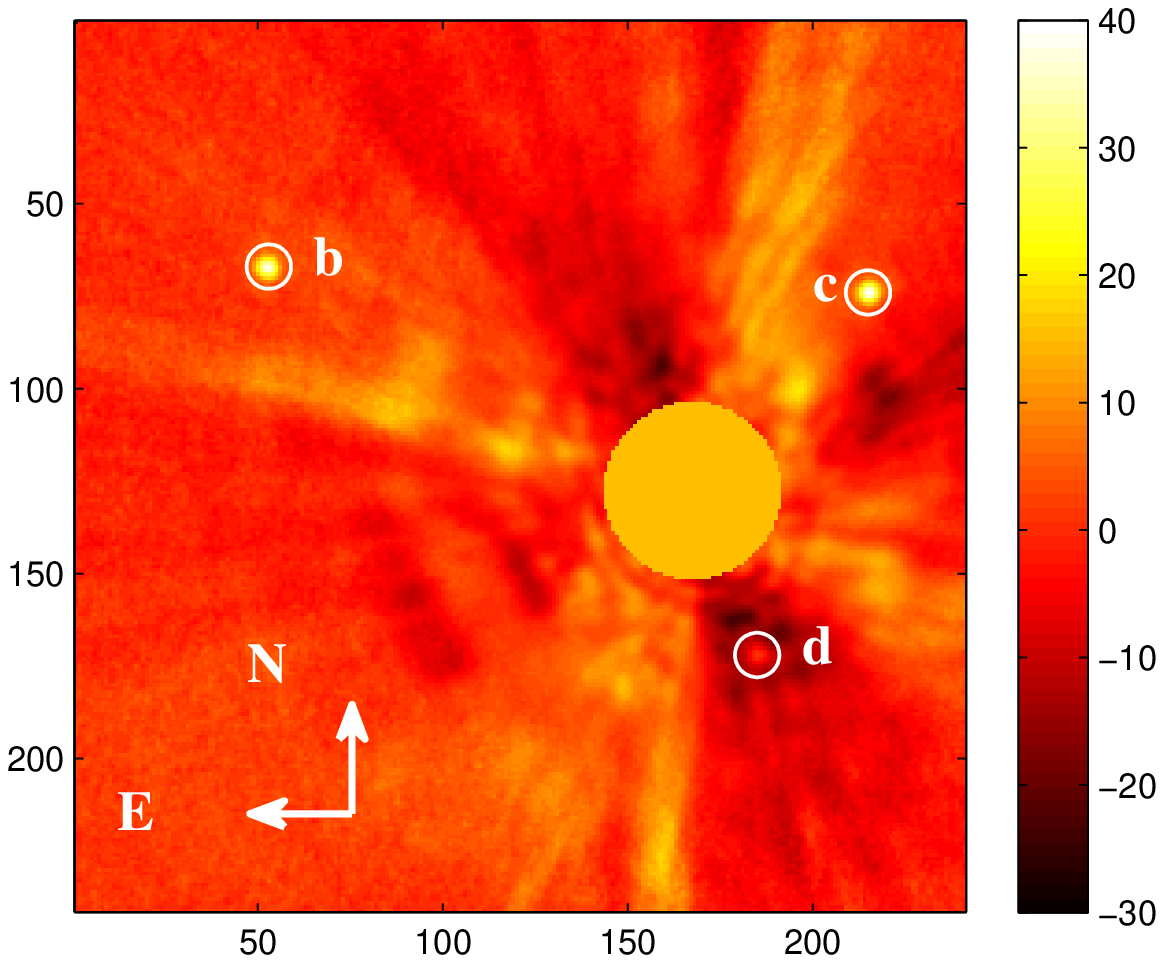}{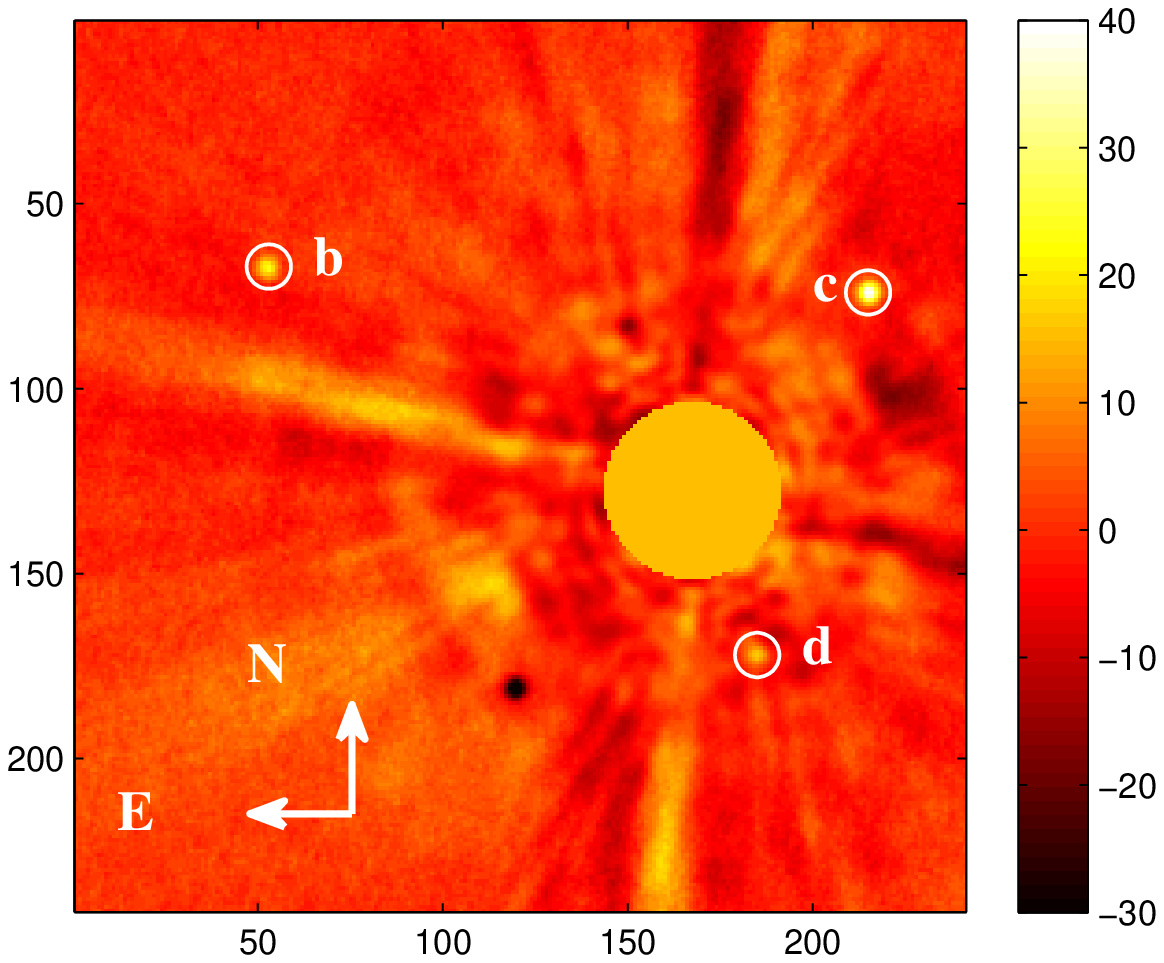}
\caption{The reduced HR 8799 images: by ADI algorithm (left panel) and IRS (right panel). These two images has been adjusted to the same contrast range. See the electronic edition of the journal for a color version of this figure.\label{fig5}}
\end{figure}


\begin{figure}
\epsscale{1.2}
\plottwo{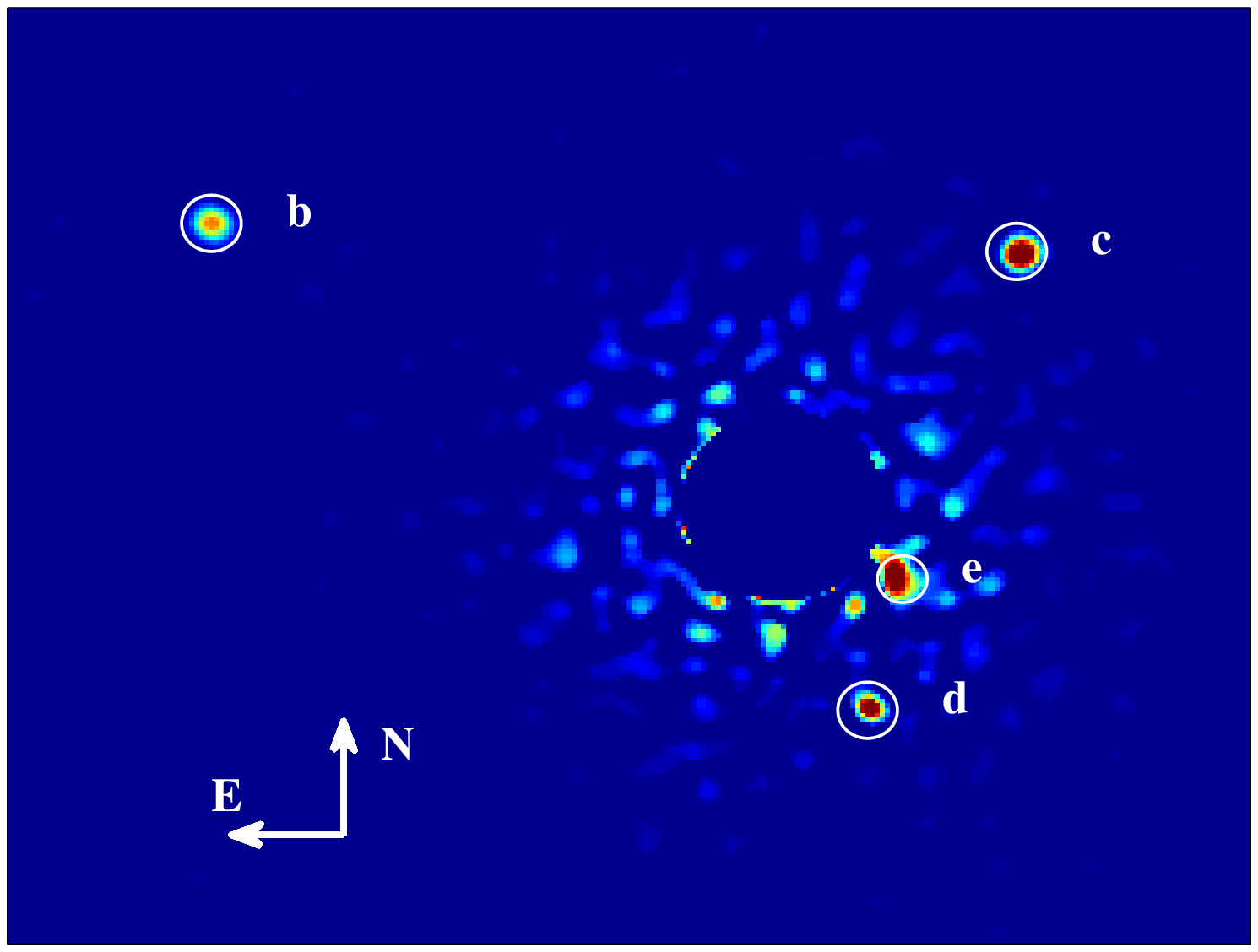}{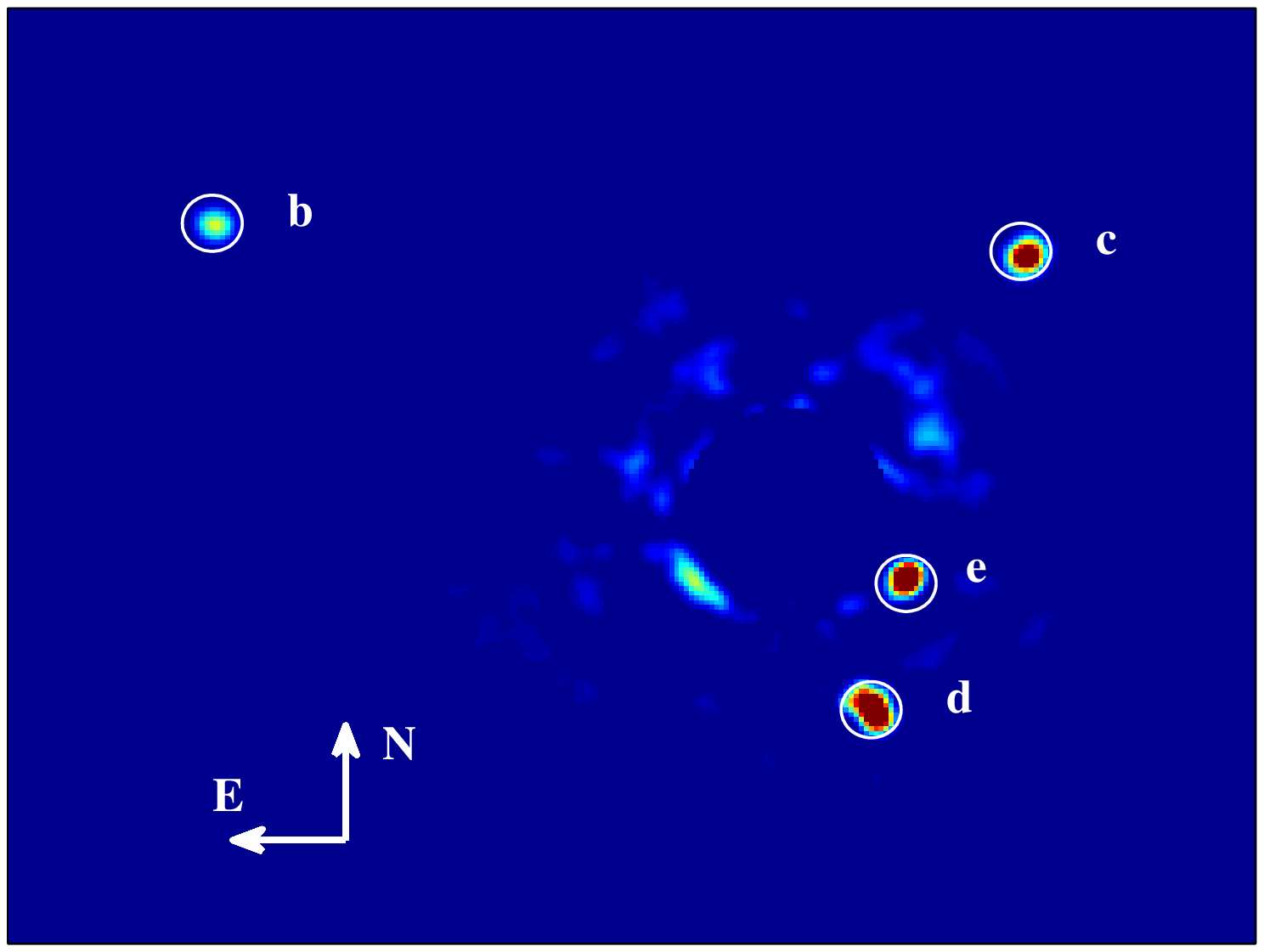}
\caption{The reduced HR 8799 images: by LOCI (left panel) and OIRS algorithms (right panel). A gauss filter has been applied to both images. These two images have been adjusted to the same contrast range. See the electronic edition of the Journal for a color version of this figure.\label{fig6}}
\end{figure}

\begin{figure}
\epsscale{2}
\plottwo{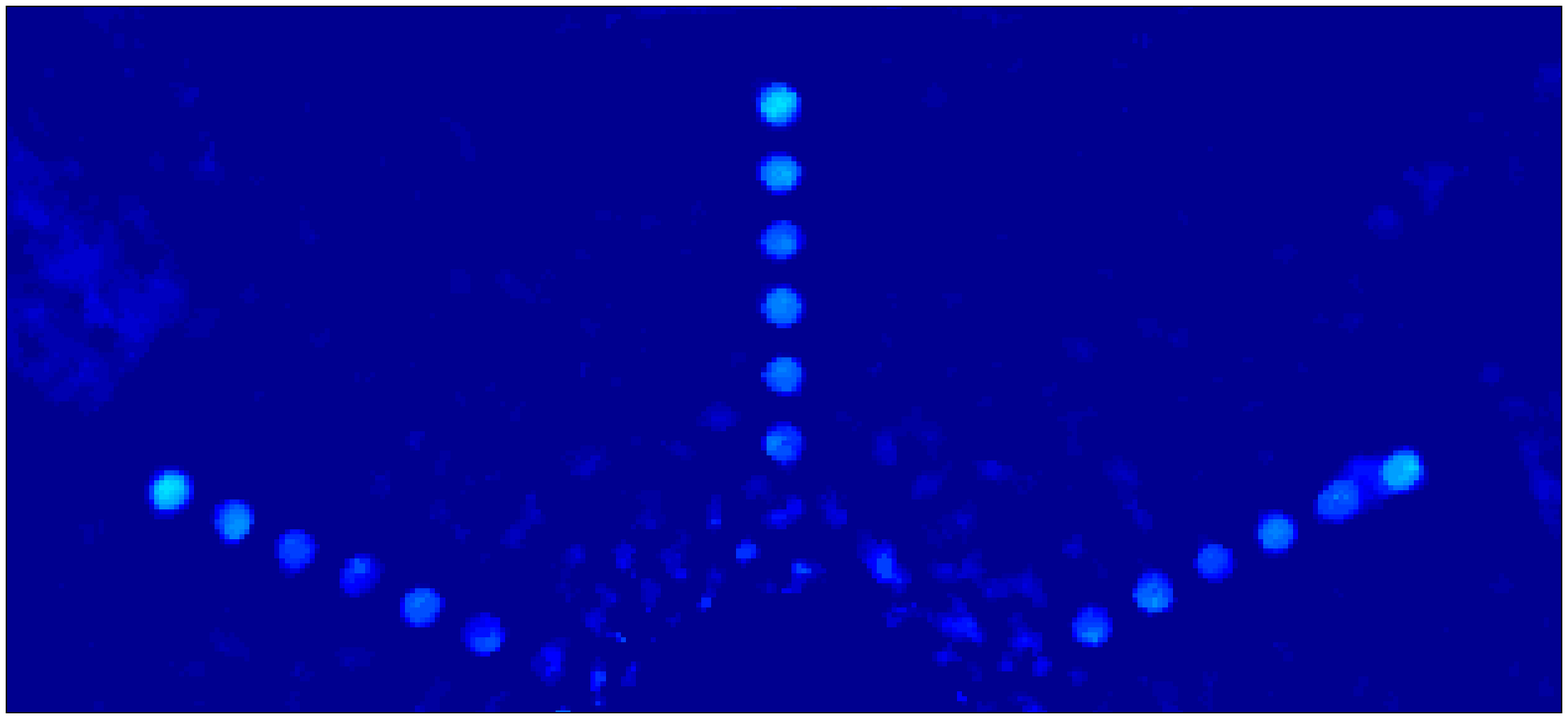}{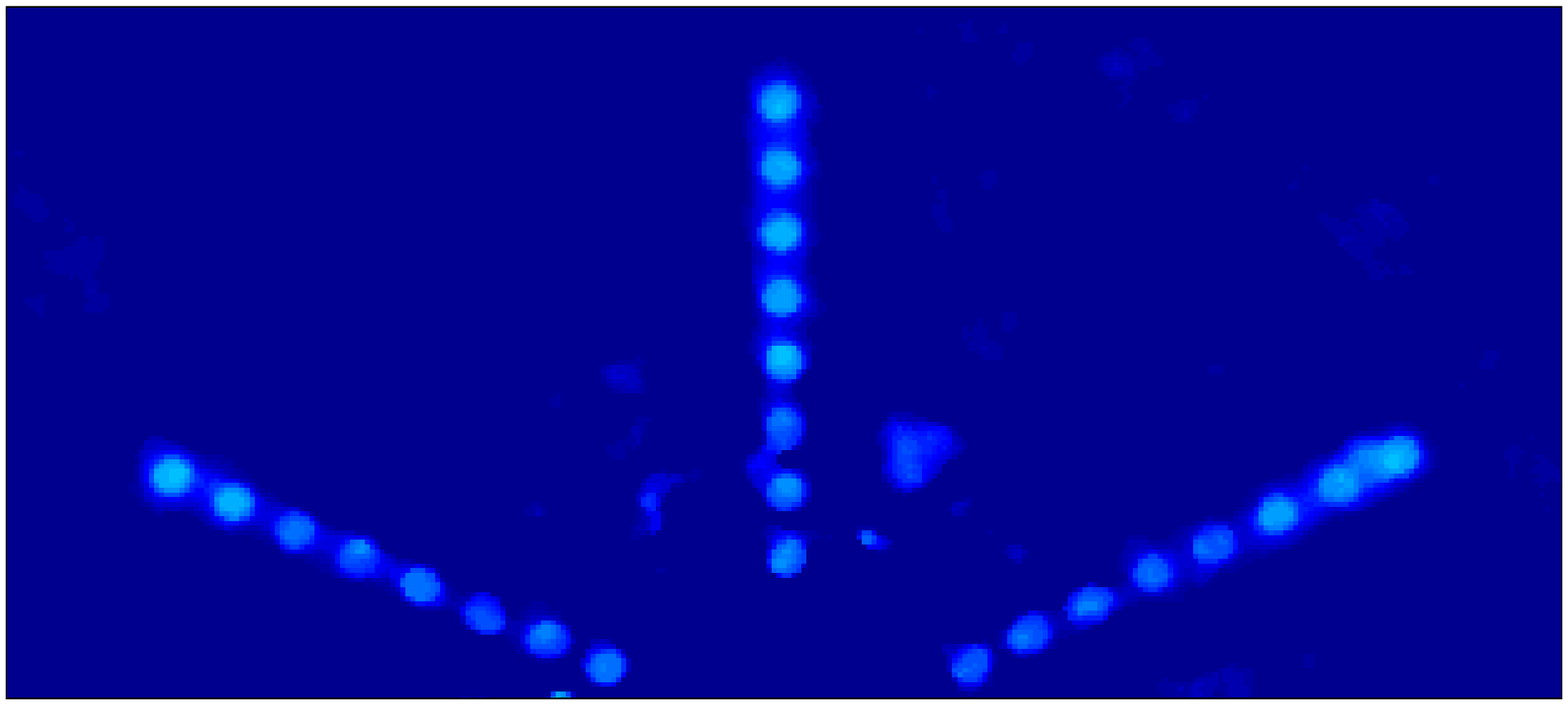}
\caption{Residual image including artificial planets by using LOCI (top) and OIRS algorithms (bottom).\label{fig7}}
\end{figure}

\begin{figure}
\epsscale{.8}
\plotone{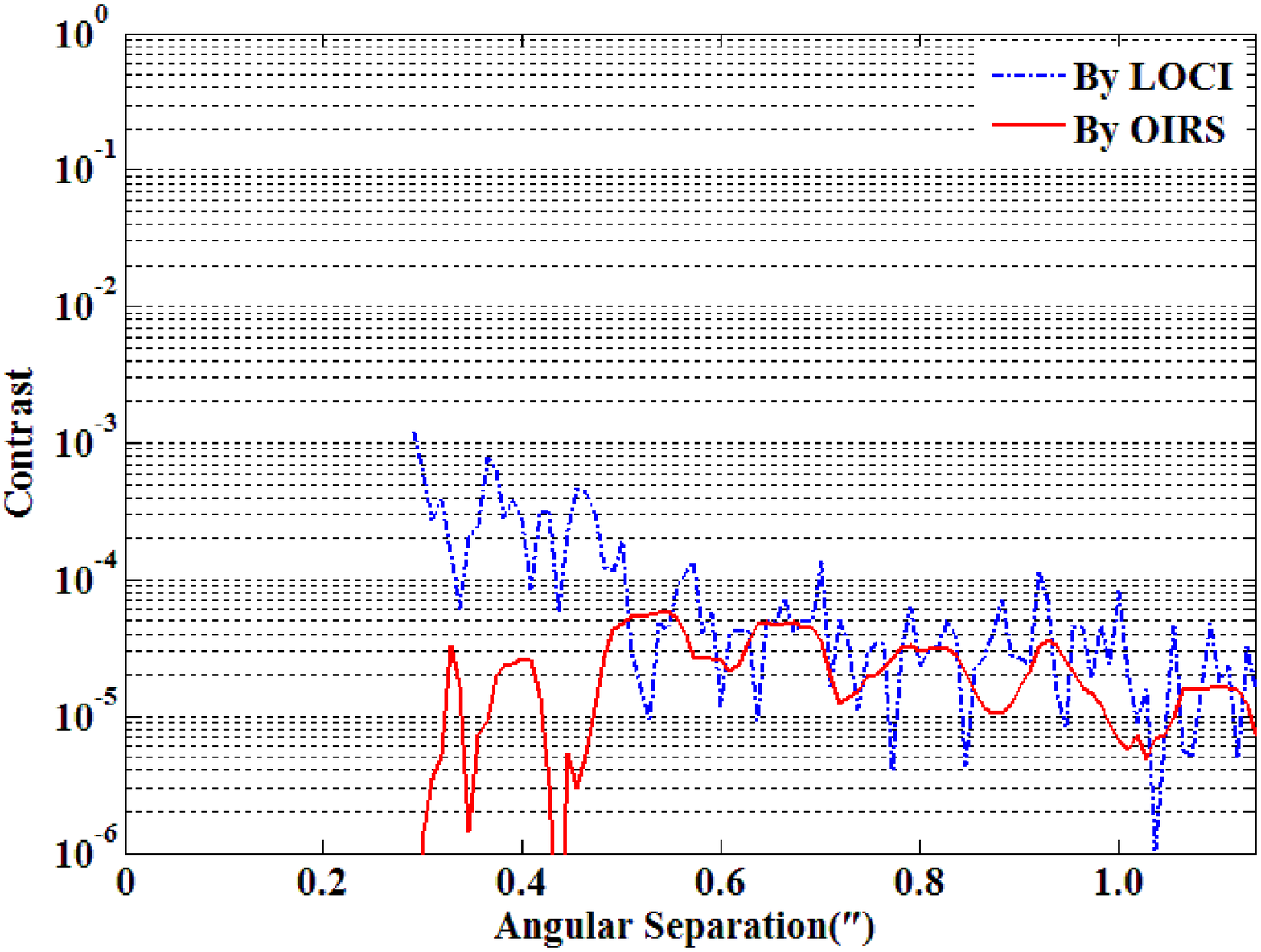}
\caption{Achievable contrast by two algorithms. See the electronic edition of the journal for a color version of this figure.\label{fig8}}
\end{figure}







\clearpage

%

\begin{deluxetable}{ccccc}
\tabletypesize{\scriptsize}
\tablecaption{Comparison of S/N\label{tbl-1}}
\tablewidth{0pt}
\tablehead{
\colhead{ } & \colhead{HR 8799 e} & \colhead{HR 8799 d} & \colhead{HR 8799 c} & \colhead{HR 8799 b}
}
\startdata
ADI & None detection &	$6 \sigma$ & $>15 \sigma$	& $>15 \sigma$ \\
IRS & None detection &	$10 \sigma$ & $>15 \sigma$	& $>15 \sigma$ \\

LOCI & $\sim 4 \sigma$ &	$10 \sigma$ & $>20 \sigma$	& $>25 \sigma$ \\
OIRS & $\sim 8 \sigma$ &	$15 \sigma$ & $>20 \sigma$	& $>25 \sigma$ \\
\enddata

\end{deluxetable}






\end{document}